 \documentclass{aa}
\input epsf 
\begin{document}

   \thesaurus{03 (11.17.3; 11.03.1)} % Buscar en alguna parte!! 
   \title{Excesses of faint galaxies around seven radio QSOs at
1.0$<z<$1.6}

   \subtitle{Clustering at high-$z$}

   \author{S.F. S\'anchez\inst{1,2}
\and
 J.I. Gonz\'alez-Serrano\inst{3}
%          \and
%          C. Ptolemy\inst{2}\fnmsep\thanks{Just to show the usage
%          of the elements in the author field}
          }

   \offprints{S\'anchez}

   \institute{
	Isaac Newton Group, Apt. de Correos 321, E-38700-La Palma,
	Spain: {\it e-mail:} sanchez@ing.iac.es
	\and
	Dept. de F\'\i sica Moderna. Facultad de Ciencias,
	 Universidad de Cantabria, E-39005 Santander, Spain
	\and
	Instituto de F\'\i sica de Cantabria(CSIC-UC).
	 Facultad de Ciencias, Universidad de Cantabria, 
         E-39005 Santander, Spain
	}

   \date{}

   \maketitle

    \begin{abstract}

We have conducted an optical study of the environments of seven
radio-loud quasars at redshifts $1<z<1.6$. In this paper we describe
deep $B$ and $R$ band images obtained for fields of $\sim$6$\times$6
arcmin around these quasars with 3$\sigma$ limiting magnitudes of
$B\sim26$ and $R\sim25.4$. We found an statistically significant
excess of {\it faint} ($B>22.5$ and $R>22.0$) galaxies on scales of
$r<$170'' and $r<$35'' around the quasars. The number of excess
galaxies, their magnitudes and the angular extension of the excesses
are compatible with clusters of galaxies at the redshift of the
QSOs. We also found that the quasars of our sample are in general {\it
not} located at the peak of the density distribution, lying in a
projected distance of about $\sim$70'' from it. This result agrees
with the scenario of a link between overdensities of galaxies around
radio sources and the origin of their radio emission.


  \end{abstract}

%
%________________________________________________________________

\section{Introduction}

Evidence has been accumulating over the last decade that there is a
link between quasar activity and the environment of the host galaxy
(e.g., Stockton 1982; Yee \& Green 1984, 1987; Gehren et al. 1984;
Hutchings et al. 1984; Hintzen 1984; Yee 1987;
\cite{eyg91}; \cite{hint91}; \cite{yg93}; \cite{fis96};\cite{yam97};
\cite{hgc98}; \cite{hg98}). In particular, the galaxy-galaxy
interaction appears to play a substantial role as the triggering or
fuelling mechanism of the nuclear activity. The large fraction of host
galaxies (HG) of radio-loud quasars undergoing tidal
interactions/merging processes (e.g. Disney et al. 1995; Hutchings \&
Neff 1997), the high absolute magnitudes of these galaxies
(e.g. Carballo et al. 1998, and references therein), and the velocity
distribution of galaxies around the QSOs (Heckman et al. 1984;
\cite{eyg91}) lend support to this scenario (\cite{yg93}).

Furthermore, Yee \& Green (1987), Ellingson et al. (1991) and Yee \&
Ellingson (1993) presented strong evidence of a link between radio
emission and clustering of galaxies around QSOs. They found that there
are significant differences between the environments of radio-loud and
radio-quiet quasars. While radio-loud quasars are associated with
clusters of galaxies of Abell richness class 0-2, radio-quiet quasars
inhabit environments of Abell richness class poorer than 0.

These results are based on searches of overdensities around
low-redshift quasars ($z\le$0.7). At higher redshifts the number of
studies are scarce, or they are focused in one or a few objects
(Hintzen et al. 1991; Boyle \& Couch 1993, Hutchings et al. 1993,
Yamada et al. 1997, Hall \& Green 1998). It seems that the dichotomy
of radio-loud/radio-quiet environment extrapolates to higher
redshifts. Hintzen et al. (1991) reported on the observation of 16
fields around radio-loud QSOs with $0.9\le z\le 1.5$. Their sample was
selected as radio-loud quasars from the V\'eron-Cetty \& V\'eron
(1984) catalog. They found a significant excess of galaxies within a
radius of 15 arcsec around the quasars. However, Boyle \& Couch (1993)
found no excess of galaxies around their sample of 27 radio-quiet
quasars, at the same redshift.

Hall et al. (1998) and Hall \& Green (1998) have recently reported on
an excess of {\it faint} galaxies ($K\ge$19) around a sample of 31
radio-loud quasars, covering a redshift range $z$=1-2. Although the
sample was observed at the optical and NIR wavelengths, the excess was
found in their $K$-band images. The excess occurs on two spatial
scales: a peak at $\theta<$35'' from the quasars and a smooth
component down to $\theta\sim$100'', which is the largest scale
sampled by their $K$-band images. The magnitudes and colors of the
excess galaxies are consistent with a population of early-type
galaxies at the quasar redshift. This excess is stronger for the more
radio-powerful and steep-spectrum radio-quasars, and it is consistent
with a cluster of galaxies with Abell richness class $\sim1.5\pm1.5$.

It is still an open question whether the most important factor
affecting the evolution of radio-loud QSOs in clusters is the absolute
density of the environment, the density within $\sim$0.5Mpc, or even
the conditions in the host galaxy and any galaxies interacting with it
(Hall \& Green 1998). It is important to determine if clusters around
quasars at high-$z$ are really embedded in large-scale overdensities,
or if they are young, less-peaked, smoothed clusters that extend to
large scales ($\sim$2-3Mpc). Paraphrasing Hall \& Green (1998)
``...larger areas (4'$\times$4') around our sample need to be
imaged to confirm the large-scale ($\theta\sim$100'') galaxy excess
around z=1-2 radio-loud quasars and to determine the true angular
extent of this excess...''.

In this paper we report on the results obtained from deep CCD imaging
survey of the fields around 7 radio-quasars selected from the B3-VLA
quasar sample (Vigotti et al. 1997), in the redshift range
$1<z<1.6$. The quasars are steep-spectrum radio-sources and they have
been selected as the brightest ones of this sample at this $z$ range
($\langle z\rangle \sim 1.4$), using the available optical information
when the selection was made (Vigotti et al. 1997). New optical
observations show that it is not true any longer (Carballo et
al. 1999), but they are still on the range of the brightest ones, having
$\langle M_{\rm B}\rangle\sim -26.6$. The size of the images
(6'$\times$6') will allow us to sample the density of galaxies to
scales down to $\sim$190''. In Sect. 2 we describe the observations
and data analysis. In Sect. 3 we analyse the possible projected
clusters around the QSOs, describing some of their properties. In
Sect. 4 we compare with previous results, and discuss about possible
interpretations. The conclusions are presented in Sect.
5. Throughout this paper we adopt $H_0$=50
km s$^{-1}$Mpc$^{-1}$ and q$_0$=0.5 .

\section{Observation and data processing}

Observations were carried out during the night of 1997 March 13 at the
Cassegrain focus of the 2.2m Telescope at Calar Alto (Spain). $B$ and
$R$ standard Johnson's filters were used and the pixel scale was 0.533
arcsec$\cdot$pixel$^{-1}$. The field of view corresponds to $\sim$6$'
\times 6'$ in all the images. The seeing was $\sim$1.6'', and variations
along the night were lower than 0.2''.

An standard observing procedure was used. CCD bias were measured
taking 0 s. exposure time images. Sky flat-fields for each filter were
obtained during twilight. Dome-flats were also obtained in order to
check the sky flat-fields stability; however, sky flat-fields have
been found to work properly and to produce better flat-fielding. In
order to reduce the cosmic ray events, and obtain a proper cleaning of
them, the following observing procedure was applied: two frames of 900
seconds exposure time at the $R$ band and three frames (two of 900
seconds and another one of 600 seconds) at the $B$ band were obtained
for each QSO.

The individual frames were reduced using standard tasks in
IRAF$^1$\footnotetext[1]{IRAF is distributed by the NOAO, which is
operated by AURA, Inc., under contract to the NSF} package. First, the
corresponding bias was subtracted for each image. A flat-field image
was obtained for each filter using the average of the sky flats with
higher signal-to-noise and without saturated pixels. The images were
corrected using these flat-fields, and the error produced by this
correction was lower than 0.8\%. Cosmic ray events in each frame were
detected and replaced by the average of the neighbours using the
COSMICRAYS task on IRAF packages.

The final images, with 1800 and 2400 seconds exposure time for the $R$
and $B$ band respectively , were obtained by coadding the reduced
individual frames for each object and filter.

\subsection{Photometric calibration}

Flux calibration for each night and filter was carried out by a
standard procedure using 28 Landolt faint photometric standard stars
(Landolt 1992, and references therein), that were observed for each
filter along the night. The extinction coefficient and {\it
zero-point} were obtained by a least-square fitting between the
instrumental magnitude ($-2.5\cdot {\rm log}_{\rm 10}(N_c/t)$, where $N_c$
is the number of counts), and the air-mass ($X$). The night was
photometric, and the $rms$ of these least-square fits were 0.086 and
0.027 for the $B$ and $R$ band respectively.

The QSO magnitudes were measured on the images using circular
apertures centred at the emission peak. The aperture radius was fixed
for the seven QSOs to 6.4'', and it was set to the value at which the
intensity of the object reached the background for the QSO with the
larger seeing image. Table 1 lists a summary of these results,
including some useful radio properties of the quasars and their
redshifts.  Carballo et al. (1999) report on the optical photometry
of 73 B3-VLA quasars. They observed the seven objects of our sample,
in both the $B$ and $R$ band, and quote magnitudes consistent with
those reported here, within an error of $\sim$0.1 mag.

%__________________________________________________ One column table
   \begin{table}
      \caption[]{QSOs sample}
         \label{table1}

      \[
         \begin{tabular}{lrrrrr}
            \hline
            \noalign{\smallskip}
% Cabezera!!
{B3 Name}
&{$z$}
&{$\alpha_{\rm 408}^{\rm 1460}$}
&{$P_{\rm 408}$ $^{a}$}
&{ $B$ }
&{ $R$ }\\
%            \noalign{\smallskip}
            \hline
            \noalign{\smallskip}
%Tabla!!
0740+380C&1.063&-1.29& 7.88&18.36$\pm$0.09&17.65$\pm$0.03\\
0926+388 &1.630&-1.09& 1.84&19.68$\pm$0.09&19.22$\pm$0.03\\
1123+395 &1.470&-0.82& 1.05&18.54$\pm$0.09&17.96$\pm$0.03\\
1148+387 &1.303&-0.94& 4.08&17.36$\pm$0.09&16.74$\pm$0.03\\
1206+439B&1.400&-0.85&14.90&18.41$\pm$0.09&17.39$\pm$0.03\\
1315+396 &1.560&-0.49& 3.87&18.70$\pm$0.09&18.28$\pm$0.03\\
1435+383 &1.600&-0.85& 1.91&18.23$\pm$0.09&17.83$\pm$0.03\\
            \noalign{\smallskip}
            \hline
         \end{tabular}
      \]
\begin{list}{}{}
\item[$^{a}$] in $10^{28}$ W Hz$^{-1}$
\end{list}
   \end{table}

\subsection{Data processing and number counts}

We searched for galaxies in the frames using the SExtractor package
(\cite{ber96}). This program detects and deblends objects using an
isophotal threshold above the local sky and produces a catalogue of
their properties.  This catalogue includes information of position,
shape, profile, magnitude and type, for each detected object. Objects
are classified using a neural network, which takes as input parameters
8 isophotal areas, the peak intensity and the FWHM (full width at half
maximum) of each object.  The $seeing$, defined as the FWHM of the
stars, is another {\it input} parameter.  This network determines a
``star-like-index'', which is, by definition, 0 for the galaxies and 1
for the stars. In practice, a value of $\sim 0.7$ works fine to
separate stars and galaxies.

We selected a detection threshold of 5 connected pixels
($\sim$1.4 arcsec$^2$) above 2$\sigma_{\rm sky}$ level per pixel, which
guarantees that $all$ the detected objects have signal-to-noise larger
than 4.5. This detection criterium is similar to those reported in
previous works. For instance, Hall et al. (1998) used a catalog of
objects detected with a minimum area of 1.9 arcsec$^2$ and a detection
threshold of 5$\sigma$. We note here that for
an area of 1.9 arcsec$^2$ our detection threshold would be
5.2$\sigma$.
We applied SExtractor to the 7 $R$-band and 7 $B$-band images using
this detection criterium. After cleaning sources close to the
frame borders or near to large foreground galaxies, we obtained
the final catalogs for each field and filter.
We found that the mean magnitudes of the faintest detected
objects in the different fields are $B$=$25.58\pm0.35$ and
$R$=$24.80\pm0.28$. These are the 4.5$\sigma$ limiting magnitude of
our catalogs.

The $3\sigma$ limiting magnitude for 5 connected pixels of the images
can be estimated by measuring the standard deviation on the sky. This
yields mean values of $B^{3\sigma}_{\rm lim}$=26.0$\pm$0.4 and
$R^{3\sigma}_{\rm lim}$=25.4$\pm$0.3. This would imply a 4.5$\sigma$
limiting magnitude of $B^{4.5\sigma}_{\rm lim}$=25.6$\pm$0.4 and
$R^{4.5\sigma}_{\rm lim}$=24.9$\pm$0.3, which are consistent with the
magnitudes of the faintest detected objects in the catalogs.

The 3$\sigma$ limiting magnitude of the $R$-band images is similar
to values previously reported in recent works.
Hall \& Green (1998) and Hall et al. (1998) reported
a $3\sigma$ limiting magnitude of $r\sim$25.5 mag, for a minimum
detection area of 1.9arcsec$^2$. Taking into account the relation
between the Thuan-Gunn $r$-band and the Johnson $R$-band magnitudes
($r=R+0.43+0.15(B-V)$ ; Kent 1985), and assuming a range of $B-V$
colours between 0 and 2 (for typical elliptical galaxies at $0<z<2$),
the Hall \& Green $3\sigma$ limiting magnitude would be
$R\sim$24.8-25.1. These authors study
overdensities of galaxies around radio-quasars at $1<z<2$, whereas our
maximum redshift is $z\sim$1.6. Therefore we conclude that our images
are adequate to search for galaxies around QSOs at these redshifts.

In order to test the validity of our method we have made a set of
simulations. Seven simulated images have been built with the same
magnitude-to-counts relation, background and noise than the $R$-band
images of our sample. The simulated images were built using the
ARTDATA tasks in the IRAF package. In each image 1200
galaxies, uniformly spatially distributed, and with a number-counts
distribution following a power-law with a slope of 0.4, ranging from
$R$=18.0 to 25.5, were simualted. We added 750 stars to each
image, with magnitudes in the range 14 to 25 mag, and similar
spatial and flux distribution than the galaxies. We have artificially
increased the star/galaxy number ratio with respect to that observed
in our images in order to test the galaxy detection and classification
at faint flux levels.

The whole procedure was applied over these simulated images. The
 results have been compared with the input parameters of the objects. We
found that 85\% of the sources were correctly classified down to
$R\le$23.5. The percentage of objects (galaxies+stars)
misclassified grows at fainter magnitudes, being around 30\% at
$R>$23.5. The fraction of galaxies classified as stars is roughly
 similar to the fraction of stars classified as galaxies.
The input number-counts distribution of galaxies was
recovered down to $R\le$23.8 (objects detected up to
$\sim$13$\sigma$). A similar result is expected for the objects
down to $B<$24.5 mag.

\subsubsection{Number counts}

A proper determination of the {\it field} number-counts and its
standard deviation is crucial to detect excesses of galaxies and to
quantify their significances. Some authors take the statistical
significance using the poissionian standard deviation.  But, as have
been quoted (Yee \& Green 1987, and references therein; Ben\'\i tez et
al. 1995), galaxies tend to cluster, and the true standard deviation
depends not only on the number of objects (as in the poissonian case)
but also on the two-point correlation function. Therefore, the
standard deviation $\sigma(r,N)$ depends on the number of galaxies $N$
and on the angular scale $r$. This is usually expressed as
$\sigma(r,N)=\gamma (r)\cdot\sqrt{N}$. The factor $\gamma (r)$ takes
account of the angular scale, and since there is no galaxy-galaxy
correlations at any scale, it must tend to 1 at some large scale (Yee
\& Green 1987). The limit $\gamma =1$ corresponds to a poissonian
distribution.  Ellingson et al. (1991, and references therein) set
$\gamma(r)$ to 1.3 for scales of $\sim$60 arcsec, whereas Ben\'\i tez et
al. (1995) found a value of 1.5 for scales as large as $\sim$120
arcsec and Hintzen et al. (1991) assumed a poissonian distribution for
scales of $\sim$15 arcsec. We have prefered to make an internal
measurement of the factor $\gamma (r)$.

The number of {\it field} galaxies in the QSOs images was measured
directly and then compared with similar results available in the
literature. Two different, and complementary, measurements have been
made. {First}, the number of galaxies in the outer part of the images,
in an anulus from $r_{\rm in}$=170'' to $r_{\rm out}$=185'' around the QSO,
was measured in each image.  This radius corresponds to $\sim 2.5$ Mpc
at the mean redshift of the QSOs and guarantees that there are no
contaminations from possible clusters members.  The mean galaxy-counts
obtained for each QSO field and filter defines a density of field
galaxies. The measurement was repeated for different ranges of
magnitude, and we obtained a number-counts distribution. This method
has the advantage that possible contaminations from the putative
cluster members at these scales have been minimized. On the other hand
it has the disadvantage that there is only one measurement for each field,
and therefore, the standard deviation is not well determined. The
measured factor $\gamma$ with this method would correspond to
$\gamma(170'')$.

The second method consists in measuring
the number of galaxies in a grid of 25 circular, not
overlapping, areas of $35$ arcsec radius ($1.1$ arcmin$^2$).
This angular scale corresponds to $\sim$0.5 Mpc at the mean $z$ of the
quasars. To minimize contamination from possible clusters, only areas
at distances larger than 70 arcsec from the QSOs were considered.
This
method has the adventage that now we have 16 measurements of the number
of {\it field}  galaxies for each image, resulting in a total of 112
measurements for each band. This provides us with a well determined
$\gamma(35'')$ factor. The major caveat
is that it is not possible to guarantee that the areas
are not contaminated by cluster members. Recalling that
$\gamma(r)\neq 1$ is due to galaxy-galaxy correlations,
any clustering would produce an increase of this factor. Therefore,
any contamination by cluster members tends to increase the measured
value of $\gamma(35'')$ and to reduce the significance of the possible
excess.

Table 2 lists the number of measured field galaxies (we will refer
them as {\it expected} galaxies throughout this paper, $N_{\rm exp}$)
within an area of $\sim$1 arcmin$^2$ (a circular area of $r<35$'') at
different magnitude ranges, using the two above refereed methods. We
also list the computed $N_{\rm exp}$ taken all detected galaxies into
account ($B_{\rm all}$ and $R_{\rm all}$ in Table 2). We have re-normalized the
number of expected galaxies and its standard deviation to the same
area assumming that the distribution of $field$ galaxies is uniform
and that the deviation follows the above refereed law
($\sigma$=$\gamma(r) \sqrt{N_{\rm exp}}$). Table 2 also lists the
corresponding estimates of the $\gamma$ factor at both scales.

%__________________________________________________ One column table
   \begin{table}
      \caption[]{ Number of expected galaxies within 1.07arcmin$^2$ ($r<$35'')}
         \label{table2}

      \[
         \begin{tabular}{lrrrr}
            \hline
            \noalign{\smallskip}
% Cabezera!!
{Mag. range}
&{N$_{\rm exp}$ $^{a}$}
&{$\sigma$/$\sqrt{N_{\rm exp}}$ $^{a}$}
&{N$_{\rm exp}$ $^{b}$}
&{$\sigma$/$\sqrt{N_{\rm exp}}$ $^{b}$}
\\
            \noalign{\smallskip}
            \hline
            \noalign{\smallskip}
%Tabla!!
$B$ {\it all}&5.88$\pm$2.29&0.95&6.03$\pm$3.27&1.33$\pm$0.26\\
$B<23.5$&2.61$\pm$3.57&2.21&2.70$\pm$1.62&0.99$\pm$0.20\\
$B<22.5$&1.15$\pm$1.57&1.47&1.01$\pm$1.10&1.10$\pm$0.19\\
$B<21.5$&0.28$\pm$0.50&0.95&0.35$\pm$0.59&1.00$\pm$0.13\\
$B<20.5$&0.25$\pm$0.42&0.84&0.17$\pm$0.40&1.00$\pm$0.19\\
$B<19.5$&0.23$\pm$0.41&0.84&0.07$\pm$0.25&0.95$\pm$0.27\\
\hline
$R$ {\it all}&8.57$\pm$3.78&1.30&9.09$\pm$3.32&1.10$\pm$0.17\\
$R<23$&5.08$\pm$5.63&2.50&5.68$\pm$2.33&0.98$\pm$0.26\\
$R<22$&2.52$\pm$6.33&3.99&2.54$\pm$1.80&1.13$\pm$0.16\\
$R<21$&1.19$\pm$3.63&3.33&1.01$\pm$1.12&1.11$\pm$0.21\\
$R<20$&0.64$\pm$1.65&2.06&0.42$\pm$0.75&1.16$\pm$0.22\\
$R<19$&0.37$\pm$0.61&1.01&0.22$\pm$0.51&1.09$\pm$0.19\\
            \noalign{\smallskip}
            \hline
         \end{tabular}
      \]
\begin{list}{}{}

\item[$^{a}$]Measured in an anulus of $r_{\rm in}$=170'' and $r_{\rm out}$=185''around the QSOs
 
\item[$^{b}$] Measured in 16 circular areas of $r<$35''

\end{list}
   \end{table}

Apparently, there is no significant differences
between both measurements, at any band and magnitude range. As
expected, the first method gives a lower value of the total expected
galaxies (since there is little or no contamination from the cluster),
while the second method provides a better value of the standard
deviation. The mean value of $\gamma(r)$ is 1.79$\pm$1.05, with the
first method, and 1.08$\pm$0.11, with the second one. Both values can
not be simply compared, since they have been obtained at different
angular scales, but the larger uncertainty of the first one is clear.
If $\gamma$ is 1.08 at $r\sim$35'', at larger angular scales its value
must be between 1.08 and 1 (the poissonian case). Therefore we will use
the number counts derived from the first method and a standard
deviation of $1.08\sqrt{N_{\rm exp}}$ for scales larger than or equal to
35''. This value of the $\gamma$ factor at these scales is slightly
different from that quoted by Ellingson et al. (1991) who estimated
$\gamma \sim 1.3$. We will be comparing our results with those
obtained using this value.

As we mention above, this method guarantees little contamination from
possible cluster members, but it cannot prevent contamination from
large-scale structures. Postman et al. (1998) have noticed that such
structures could expand to distances as large as $\sim$6 Mpc (in our
adopted cosmology). It is not possible to avoid this problem with
present data, since it needs a measurement over several large-scale
structure correlation lengths to extimate the 'true' number-counts.
As in the case of 'cluster members', any
large-scale structure contamination will increase the 'real' {\it expected}
number counts. Therefore, any overdensity found would be more likely a
lower limit, to the extent that the mean density derived from our
method is slightly biased upward by the outer extent of the correlated
large-scale structure.

Fig. 1 shows the galaxy number-magnitude relation or {\it
number-counts} distribution for the $R$ and $B$ bands. The slopes of
the distributions are $\sim$0.37-0.36 (within the magnitude range
19.0$<R<$23.0) and $\sim$0.49-0.40 (within the magnitude range
19.5$<B<$23.5) respectively.  The 100\% completeness magnitudes,
defined as the magnitude were the number counts drops from a simple
power-law, are $R\sim$23 and $B\sim$23.7. These completeness
magnitudes correspond to $\sim$30$\sigma$ detections, i.e.
photometric errors near to the calibration errors ($\sim$0.03
mag). Both the expected number of galaxies and the slopes obtained are
compatible with the results of other authors down to this completeness
magnitudes (Metcalfe et al. 1991, Ellingson et al. 1991, Hintzen et
al. 1991, \cite{BW93}, Ben\'\i tez et al. 1995), with variations smaller
than $\sim$20\%. These small offsets in the number-counts quoted by
different authors are mainly due to the different quality of the
images (e.g., different seeing and photometric conditions), small
differences in the filter transmissions and the use of different
methods to detect and classify the objects (e.g., Hintzen et al. 1991,
Kron 1980).

% Figure 1
%                                                One column figure
%----------------------------------------------------------- S_vib

\begin{figure}
\vspace{0cm}
\hspace{0cm}\epsfxsize=8.8cm \epsfbox{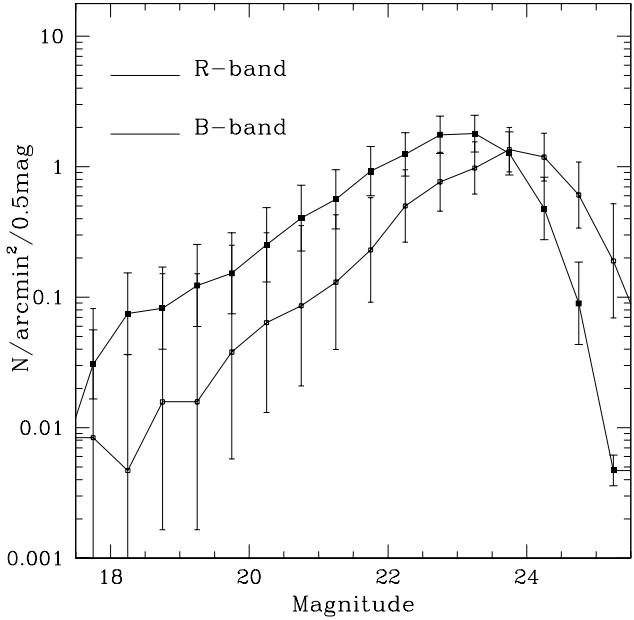}
\vspace{0cm}
       \caption[figs1.ps]{The $R$ and $B$-band galaxy number-counts
derived from the QSO images.}
	\label{Figure 1}
\end{figure}
%
%______________________________________________________________

\section{Projected clustering around the QSOs}

Since it is not known whether galaxies in the fields of
the QSOs are physically associated with them or not, it is not clear
which are the optimal magnitude range and angular scale to search for
possible excesses.
Therefore, we have preferred to use all the available
information, measuring the number of galaxies at two different scales
around the QSOs, $r_1<$170'' and $r_2<$35'', and at different
magnitude ranges. The number of galaxies found around each quasar for
each filter band has been added and later compared with the expected
number of galaxies. The number of expected galaxies when adding seven
fields was obtained using the values listed in Table 2, and were corrected for
the corresponding areas and number of fields:

$$N_{\rm exp}^{7 fields}(r_{\rm i})=7\cdot (r_{\rm i}/35'')^2
\cdot N_{\rm exp} $$

\noindent
and the corresponding standard deviation is

$$ \sigma_{N_{\rm exp}}^{7
fields}(r_{\rm i})=\gamma(r_{\rm i})\cdot\sqrt{N_{\rm exp}^{7
fields}(r_{\rm i} )}$$

\noindent
where $r_i$ is either $r_1$ or $r_2$.

We have splitted the sample in bins of 1 magnitude down to the above
defined ``100\% completeness magnitude'' and an additional bin for
galaxies fainter than this magnitude. It is important to note here
that all the galaxies, even galaxies fainter than the 100\%
completeness magnitude ($R>23$ mag, $B>23.7$ mag), were detected up to the
4.5$\sigma$ level. Moreover, the number of expected galaxies was measured
in the same images using the same detection/classification procedure
as the ``excess'' galaxies, and, therefore, they are affected by the
same incompleteness effects.
We define the relative excess (${N-N_{\rm exp}}\over{N_{\rm exp}}$) as $\eta$, and
the significance of the excess
(${N-N_{\rm exp}}\over{\gamma\sqrt{N_{\rm exp}}}$) as $n\sigma_{\gamma}$.

Table 3 lists the number of detected galaxies at both scales, the number of
expected galaxies and the significance of the excess for each band. The
significance was computed using $\gamma=$1.08, and also $\gamma=$1.3 by
comparison. These values are listed for any range of magnitude, indicated
by $all$, and for the above defined bins of magnitude. We found that there
is a significant excess of galaxies around the QSOs for both angular scales
and filter bands. The significance of the excess ranges from 3.56$\sigma$
for the $B$ band and the largest angular scale, to 4.87$\sigma$ for the $B$
band and the smallest angular scale. The excesses are still significant if
we use $\gamma=1.3$. The overdensity is detected both in the $B$ and $R$
bands at a $3 - 4\sigma$ level, and, therefore, the significance of the
detection is larger than if it were detected only in one of the bands. A
quadratic propagation of the significance of the excesses in both
bands shows that the detection is significant up to 4.60$\sigma$
($B$+$R$ {\it all} in Table 3).

%__________________________________________________ One column table
   \begin{table*}
      \caption[]{Number of galaxies within $r<$170'' and $r<$35'' from all the fields}
         \label{table3}

      \[
         \begin{tabular}{lrrrrrrrr}
            \hline
            \noalign{\smallskip}
% Cabezera!!
{Mag. range}
&{N$_{\rm tot}$ }
&{N$_{\rm exp}$ }
&{$n\sigma_{1.3}$}
&{$n\sigma_{1.08}$}
&{N$_{\rm tot}$ }
&{N$_{\rm exp}$ }
&{$n\sigma_{1.3}$}
&{$n\sigma_{1.08}$}\\
%            \noalign{\smallskip}
            \hline
            \noalign{\smallskip}
%Tabla!!
 & & &$r<$170''& & & &$r<$35''& \\
\cline{3-5}
\cline{7-9}
$B$ {\it all}&1080& 960.82&2.96&3.56&  75& 41.18&4.05&4.87\\
$R$ {\it all}&1569&1398.60&3.51&4.23&  97& 59.99&3.67&3.83\\
$B$+$R$ {\it all}& & &4.60&5.53& & &5.49&6.20\\
\hline
$B>$23.5&588&529.9&1.94&2.34& 41&23.89&2.69&3.24\\
$22.5<B\le$23.5&304&241.92&3.07&3.70& 20& 9.80&2.51&3.02\\
$21.5<B\le$22.5&123&143.64&-1.32&-1.59&  7& 5.57&0.46&0.55\\
$20.5<B\le$21.5& 38& 31.92& 0.82& 0.99&  4& 1.37&1.47&2.09\\
\hline
$R>$23.0&604&559.44&1.45&1.74& 40&26.25&2.06&2.49\\
$22.0<R\le$23.0&524&423.36&3.76&4.53& 31&17.01&2.61&3.15\\
$21.0<R\le$22.0&253&219.24&1.75&2.11& 17& 8.82&2.12&2.55\\
$20.0<R\le$21.0&110& 90.72&1.56&1.87&  4& 3.71&0.12&0.15\\
$19.0<R\le$20.0& 47& 45.36&0.19&0.23&  2& 1.75&0.15&0.18\\
            \noalign{\smallskip}
            \hline
         \end{tabular}
      \]

   \end{table*}

 The excess presents a tendency with the magnitude range, being
significant only for the faintest magnitudes (up to $\sim$3$\sigma$
for $R>22$ and $B>22.5$), for both bands and angular scales. These
results indicate that most probably the excess is not due to
intervening galaxies, since there is {\it no} significant excess down
to $R<21$ or $B<22.5$ (see Ben\'\i tez et al. 1995). The excess galaxies
could be physically associated with the QSOs: at the mean redshift of
the QSOs ($z\sim 1.4$), galaxies with $R\sim$23 and $B\sim$23.5 would
have $M_{\rm R}\sim$-20.9 and $M_{\rm B}\sim$-18.6. A $k$+e
correction typical for an E galaxy has been applied, obtained using
Bruzual (1983) $c$-model and the GISSEL code (Bruzual \& Charlot
1993), assuming a $z_{\rm for}$=5, 1 Gyr of initial burst and a Salpeter
mass function (Salpeter 1955), with masses from 0.1 to 125
$M_{\odot}$. These absolute magnitudes are not unusual, and therefore,
the galaxies are compatible with being at the quasar redshifts
(Bromley et al. 1998; Muriel et al. 1998). Moreover, recent results
(Hall \& Green 1998) show that galaxies with $r>$23
mag could be associated with quasars at $z\sim$1.5. Taking into account
the conversion between $R$ and $r$ magnitudes, their
galaxies should have $R>$22.3 mag, being compatible with our result.

The spatial extension of the excesses is also compatible with this
interpretation.  Fig. 2 shows the radial distribution of the relative
excess ($\eta$) with respect to each quasar (in arcsec), for both the $B$
and $R$ band. The mean radial distribution using the seven fields is also
shown.
The overdensities present a peak down to
$\theta\sim$35''-40'' in five of the seven fields (at least in one of the
two bands). The peak is also visible in the mean distribution of the
relative excess. The excess drops smoothly
from this peak, and remains flat until $\theta\sim$160''-170''. If the
overdensity is due to  clusters of galaxies at the redshift of
the QSOs ($z\sim1.4$), the peak would extend to $\sim$350 kpc and the
cut-off would be at about $\sim$1500 kpc. These scales are compatible with
the physical scales of known clusters at lower redshifts (e.g., Dressler
1980, Ellingson et al. 1991).

The overdensity is produced by $\sim$20-40 galaxies in each field (for
both angular scales), which corresponds to low density clusters
 (Abell richness class between 0-1, Abell 1958
). Summarizing, the magnitudes, angular sizes and number of galaxies
of the excess indicate that most probably we are detecting clusters around
the quasars.

% Figure 2
   \begin{figure*}
\vspace{0cm}
\hspace{0cm}\epsfxsize=16cm \epsfbox{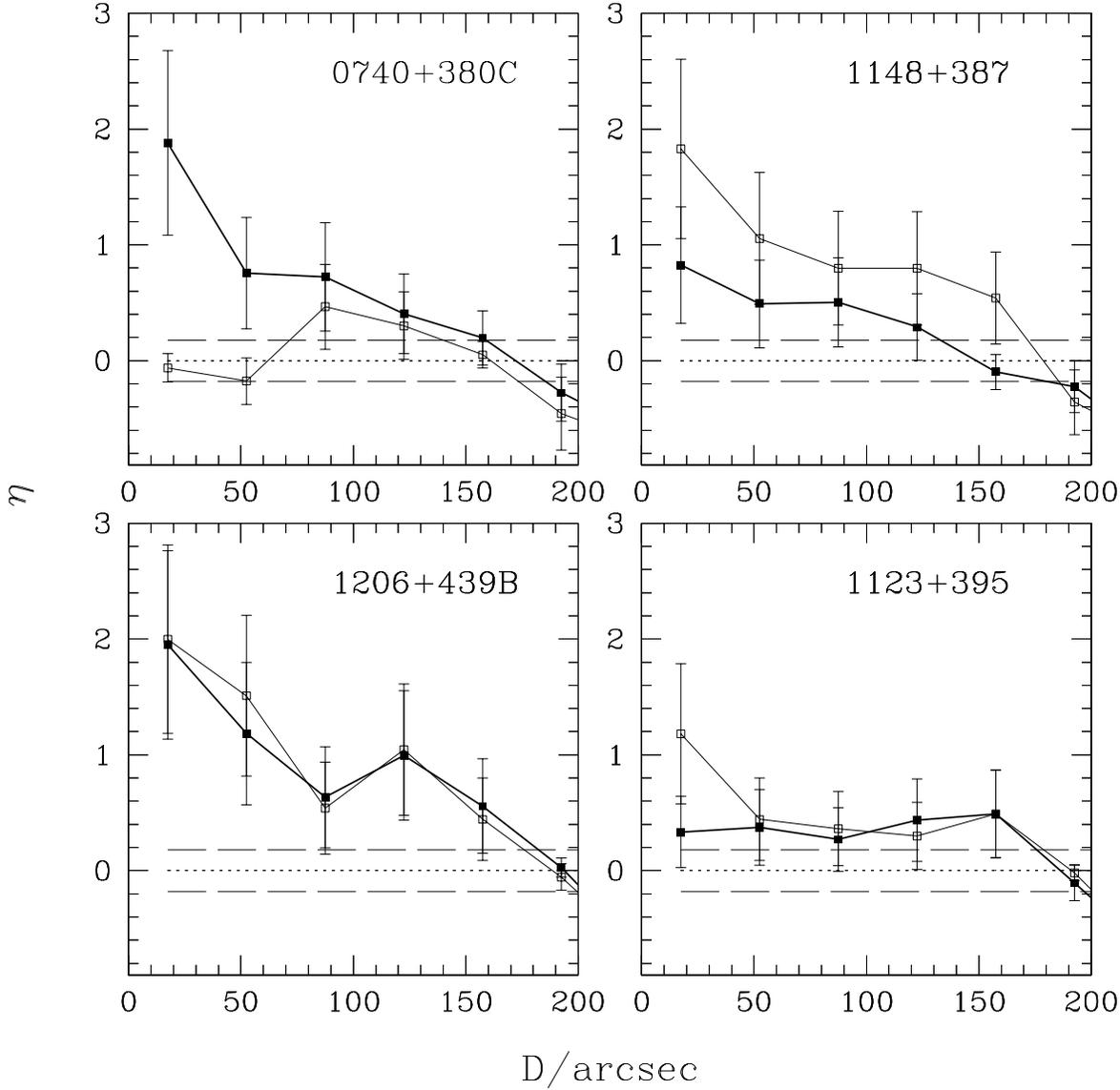}
\vspace{0cm}

\caption{Radial distribution of the relative
excess of galaxies ($\eta$) with respect to
the QSOs (in arcsec) for each quasar field and averaged over all the
fields (indicated as $mean$). The thick and thin solid lines
correspond to the objects detected in the $R$-band and $B$-band
images, respectively.}
         \label{Figure 2}%
\end{figure*}

\setcounter{figure}{1}
% Figure 2
   \begin{figure*}
\vspace{0cm}
\hspace{0cm}\epsfxsize=16cm \epsfbox{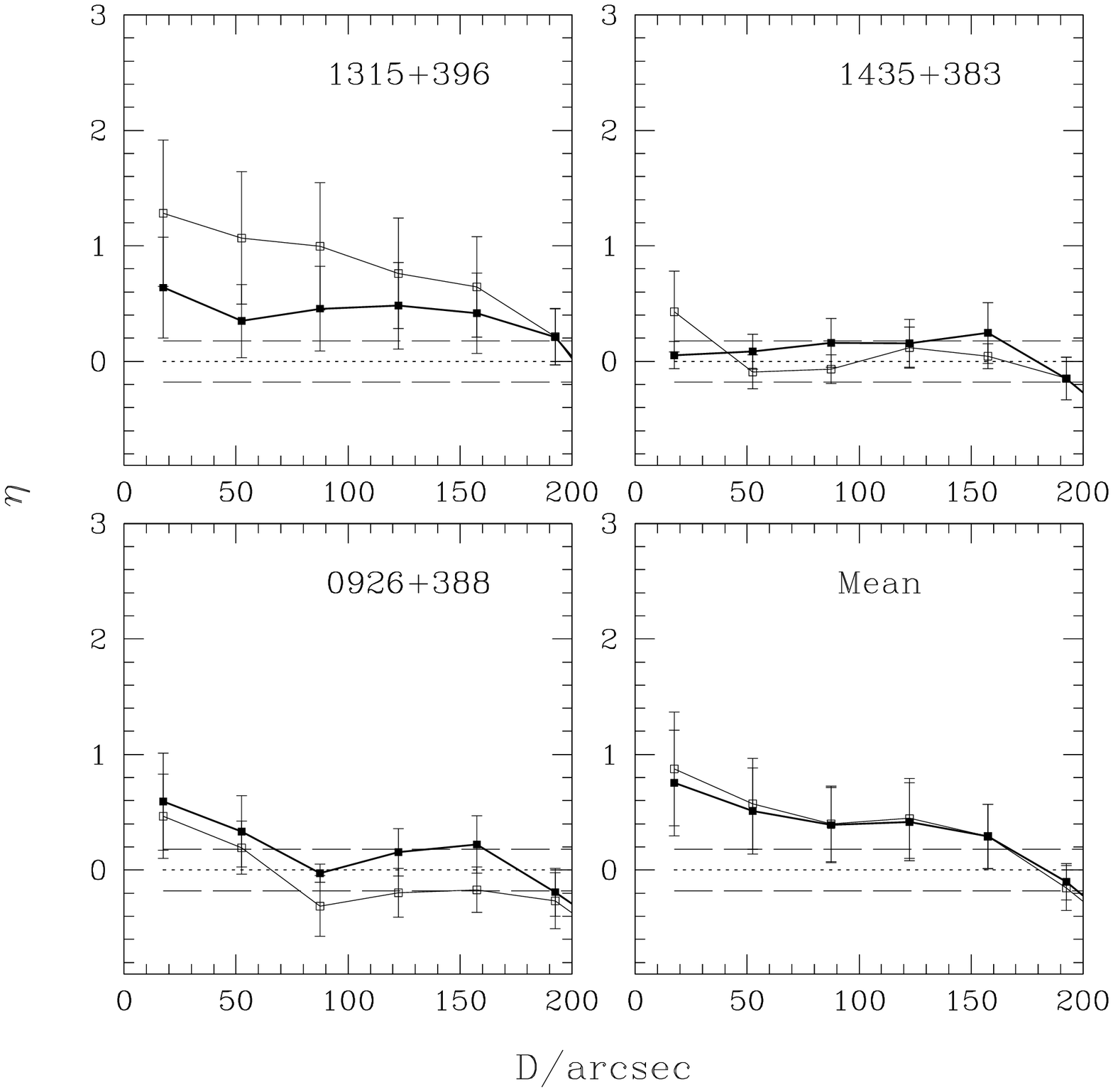}
\vspace{0cm}

\caption{{\it Continued} Radial distribution of the relative
excess of galaxies ($\eta$) with respect to
the QSOs (in arcsec) for each quasar field and averaged over all the
fields (indicated as $mean$). The thick and thin solid lines
correspond to the objects detected in the $R$-band and $B$-band
images, respectively.}%

\label{Figure 2: continued}%

    \end{figure*}

\subsection{Dependence of excesses with QSO properties}

The significance of the excess is at least $\sim$4.6$\sigma$ adding the
information from both filters and all the fields. This indicates that
the mean significance of the excess, quasar to quasar, must be
$\sim$1.7$\sigma$. It is interesting to test if the field-to-field
contribution to the total excess is uniform, or it depends on any property of
the quasars.

Table 4 lists the number of detected galaxies at $r<35$'' from the
quasars in each quasar field, together with the significance of the
excess (or defect), for each filter band. There are objects where the
excess is higher than 4.5$\sigma$, but others present unsignificant
deficits. Moreover, although mainly all the quasar fields present a
central excess, there are large differences in the spatial
distribution of the excess from quasar to quasar (see Fig. 2). This
could indicate that there is a dependence of the excess with some
property of the quasars.

In order to investigate this possibility we have studied possible
correlations between the significance of the excess at $r<35''$ and
different properties of the radio quasars. No correlation was found
between the significance of the excess, $n\sigma_{1.08}$ (of the $B$
and $R$ band and at any angular scale), and redshift, absolute
magnitude, $B-R$ colour or radio spectral index of the QSOs. Fig. 3
shows the tendency found between this significance and the radio power
of the quasars at 408MHz. The correlation coefficients between both
parameters are $r_{\rm xy}$=0.8 for the $R$ band counts, $r_{\rm xy}$=0.4 for
the $B$ band, and $r_{\rm xy}$=0.7 for the mean of both (mean without
statistic meaning). This tendency could be thought to be due to the
large contribution of the B3 1206+439B field, but, removing it, there
is still a tendency for the $R$ band counts ($r_{\rm xy}$=0.7). Since our
sample is short to be statistically significant this tendency has to
be confirmed over larger samples of radio-quasars, but it were
expected if there is a connection between radio emission and
clustering (\cite{yg93}, and references therein).

% Figure 3
%                                                One column figure
%----------------------------------------------------------- S_vib

\begin{figure}
\vspace{0cm}
\hspace{0cm}\epsfxsize=8.8cm \epsfbox{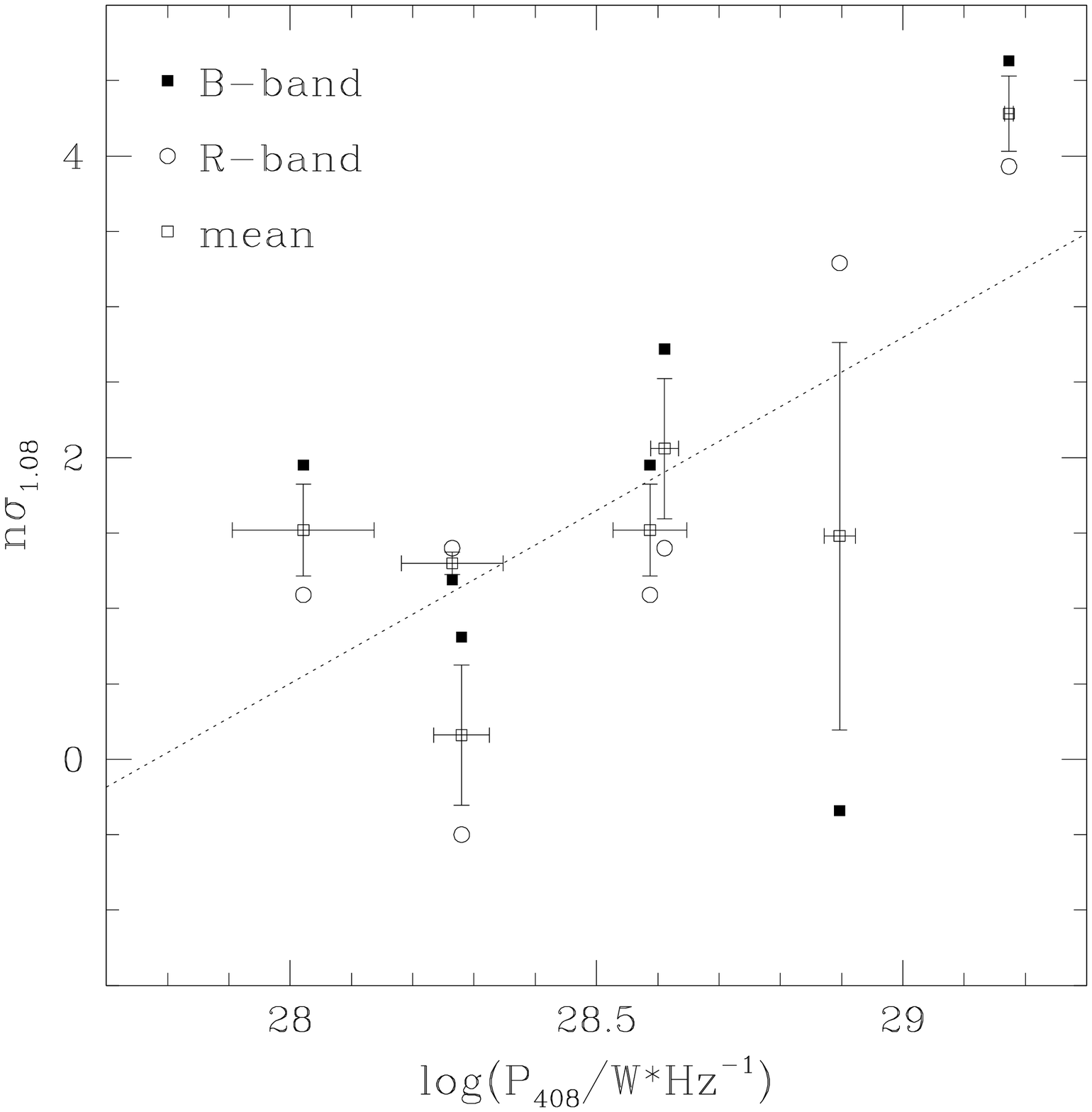}
\vspace{0cm}
       \caption[figs1.ps]{Significance of the relative excess vs.
radio power of the quasar. The dotted-line shows the result of a linear
least-square fit over the mean significance.}
	\label{Figure 1}
\end{figure}
%
%______________________________________________________________

%__________________________________________________ One column table
   \begin{table}
      \caption[]{Number of galaxies in each field}
         \label{table4}

      \[
         \begin{tabular}{lcccc}
            \hline
            \noalign{\smallskip}
% Cabezera!!
{B3 name}
&{$N_{\rm B}$}
&{$n\sigma_{1.08}$}
&{$N_{\rm R}$}
&{$n\sigma_{1.08}$}\\
%            \noalign{\smallskip}
            \hline
            \noalign{\smallskip}
%Tabla!!
 & & & $r<$35''& \\
\cline{2-5}
0740+380C&  5  &-0.34& 19& 3.29\\
1148+387 & 13  & 2.72& 13& 1.40\\
1206+395 & 18  & 4.63& 21& 3.93\\
1123+387 & 11  & 1.95& 12& 1.09\\
1315+439B& 11  & 1.95& 12& 1.09\\
1435+396 &  8  & 0.81&  7&-0.50\\
0926+383 &  9  & 1.19& 13& 1.40\\
Expected  & 5.9&     &8.6&     \\
            \noalign{\smallskip}
            \hline
         \end{tabular}
      \]

   \end{table}

\section{Comparison with other results at the same $z$ range}

Although there are numerous studies about clustering around QSOs at
redshifts lower than 0.9 (\cite{yg87}, \cite{eyg91}, \cite{yg93},
\cite{fis96}, and references therein), the number of studies decreases
at higher redshifts (Hintzen et al. 1991, Boyle \& Couch 1993,
\cite{hg98}) or they are focused in one or a few number of fields
(Hutchings et al. 1993). At low redshift appears to be clear that
radio-loud quasars inhabit groups or clusters of galaxies with an
Abell richness class ranging between 0 and 2, and radio-quiet sources
inhabit less dense groups of galaxies (e.g., \cite{eyg91}). These
results appear to reproduce at high redshift: Hintzen et
al. (1991) found a significant excess of galaxies around 16 radio-loud
QSOs with 0.9$<z<$1.5 (32 galaxies with $R<$23 within 15 arcsec when
the expectation was 19.3), while Boyle \& Couch (1993) found no
significant excesses around 27 radio-quiet QSOs in the same range of
redshifts. This dichotomy may imply a connection between clustering
and radio emission (\cite{yg93}), which would explain the tendency
between relative excess and radio-power found in our sample. Recently,
Hall et al. (1998) and Hall \& Green (1998) reported their
results on optical and NIR observations of a sample of 31 $z$=1-2
radio-loud quasars. They found a significant excess of {\it faint}
($K\ge$19) galaxies in these fields, which are compatible with being
at the quasars redshifts.

We have found  overdensities of galaxies around our radio
quasars similar to those reported by Hintzen et al. (1991) and Hall \& Green
(1998). Hintzen et al. (1991) found that their excess galaxies are
$\sim$0.5-0.8 mag brighter than first-ranked galaxies at this
redshift, but only if {\it no-evolution} is assumed. This result has
been used as an argument for the presence of gravitational lensing in
these objects (Webster 1991). However, as it has been quoted
above, the observed galaxies could be at the redshifts of the QSOs if
we take in account a small amount of evolution, like in a E-type
galaxy.

Hall \& Green (1998) found that the excess increases with power and
spectral index of radio-emission. This tendency with spectral index
was also recently reported by Mendes de Oliveira et al. (1998). Our
results (Sect. 3.1) agree with the first tendency, although we can
not test the second one due to the small baseline in radio spectral
index of our sample
($\langle\alpha_{408}^{1406}\rangle$=-0.91$\pm$0.25). Hall \& Green
(1998) also found that the excess was distributed in two different
components: a peak within $\theta<$40'' from the quasars, and a smooth
component along all the covered area. It is necessary to recall here
that their results were based on images that allowed to sample the
overdensities within $\theta<\sim$100'' from the quasars. In Fig. 2,
it is possible to see that in almost all our fields the overdensities
present a weak peak within $\theta\le$40'' from the quasar, and a
smooth component down to $\theta\sim$160''. This structure is present
in the averaged distribution (Fig.2). Hall \&
Green(1998) found that the peak is significant only in $\sim$25\% of
the fields, although the large-scale overdensity is present in all of
them. This is also seen in our sample: only 1148+387 and 1206+439B
fields present a strong peak in their central region.

Three different scenarios could explain these results: (1) the quasars
inhabit clusters that reside in large-scale overdensities of galaxies,
as has been proposed by Hall \& Green (1998), (2) they inhabit
dynamically young, still not virialized, clusters, which expand
smoothly to larger-scales (Ellingson et al. 1991 and Hall \&
Green 1998), and (3) not all the quasars are at the center of the
observed overdensities. This last possibility has not been tested
before, and it could easily explain the observed distribution. In
fact, only when the overdensity presents a clear peak in the inner
region (within $\theta<$40'') it is possible to say that the quasar is
roughly at the center of the overdensity.

In order to test this possibility we have estimated the density
distribution of galaxies of our $R$-band fields. The density
distribution has been calculated using the {\it generalized variable
kernel estimate} (Silverman 1986) using the {\it Epanechnikov kernel}
(Epanechnikov 1969). This method is an improvement of the {\it nearest
neighbour} approach, which adapts the amount of smoothing to the local
density of galaxy counts. While the most widely used estimator of the
density is based on the number of observations falling in a box of
fixed width centred at the point of interest, the kernel estimator is
inversely proportional to the size of the box needed to contain a
given number of observations ({\it k}), and also inversely
proportional to a smoothing parameter ({\it h}). The number of
observations, {\it k}, were selected to be roughly the mean density of
galaxy counts, and the smoothing parameter was arbitrarily fixed to
two points. This density estimate defines better the properties of the
distribution than the generally used (see discussion in Silverman
1986).

Fig. 4 shows a contour plot of the density distribution of galaxy
counts for each field. The first contour was plotted at the level of
the mean density, and the separation between sucessive contours is
1$\sigma$. The position of the quasars in each field was marked as a
blue star. It is clearly seen that the quasars are not at the maximum
of the distributions, hence, they are not at the ``center'' of the detected
overdensities. The distance between the quasars and the peak of the
overdensity range between $\sim$40'' and $\sim$100''.

% Figure 4
   \begin{figure*}
\vspace{0cm}
\hspace{0cm}\epsfxsize=16cm \epsfbox{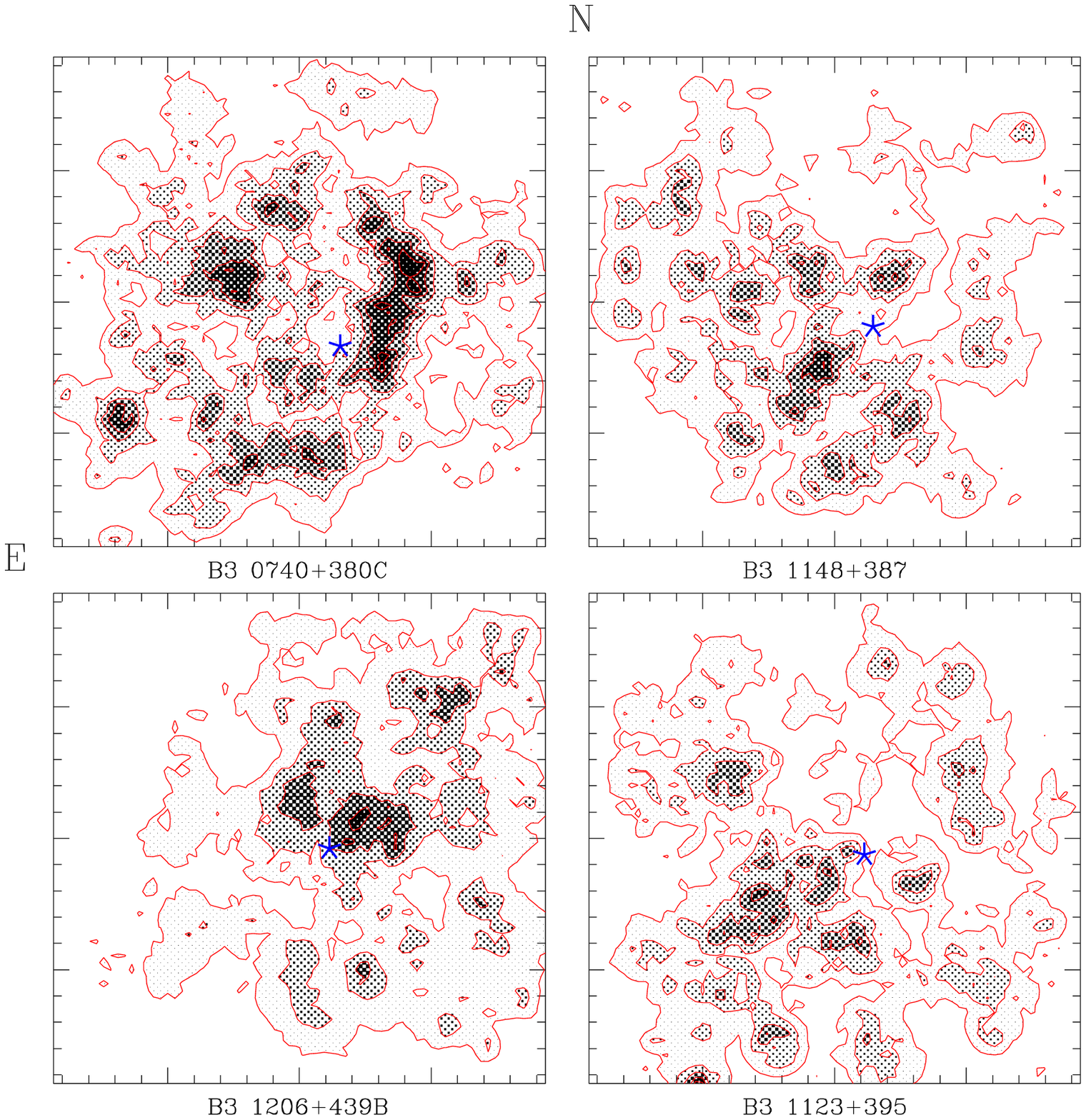}
\vspace{0cm}

\caption{Contour plots of the estimated density
distribution of galaxy counts in each field. The first contour
corresponds to the mean density, and the separation between sucessive
contours is 1$\sigma$. The position of the quasar is marked with
a blue star. The distance between small ticks is 20 arcsec.}
         \label{Figure 4}%
\end{figure*}

\setcounter{figure}{3}
% Figure 4
   \begin{figure*}
\vspace{0cm}
\hspace{0cm}\epsfxsize=16cm \epsfbox{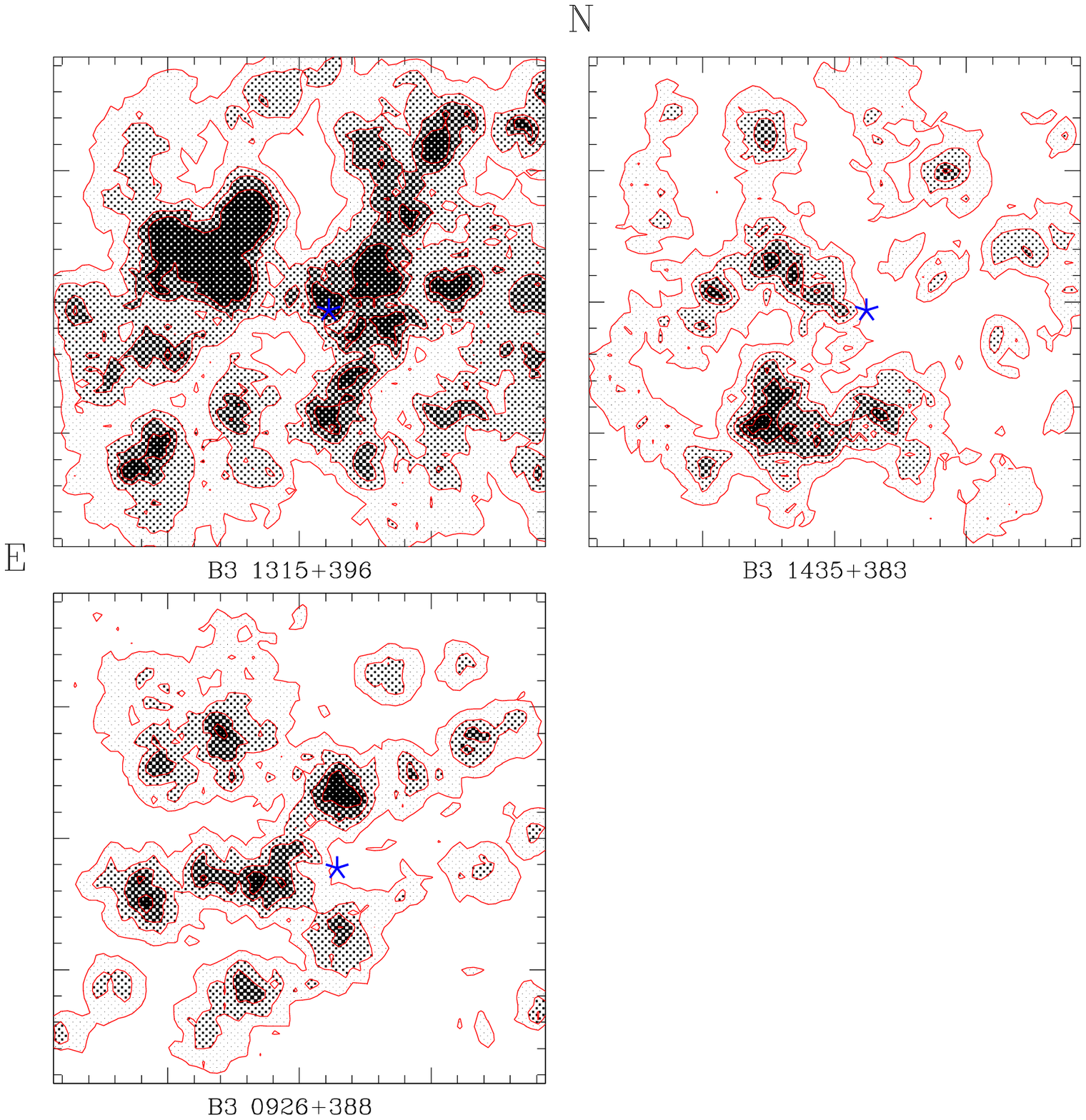}
\vspace{0cm}
\caption{{\it Continued} Contour plots of the estimated density
distribution of galaxy counts in each field. The first contour
corresponds to the mean density, and the separation between sucessive
contours is 1$\sigma$. The position of the quasar is marked with
a blue star. The distance between small ticks is 20 arcsec.}
         \label{Figure 4b}%
\end{figure*}

We have recomputed the radial distributions of the relative excess taking
as the origin the ``center'' of the observed galaxy densities. The result
is shown in Fig. 5 for all the quasar fields and for the mean, where the
position of the QSOs is indicated with an arrow.
 It is seen that these excess distributions are
more intense at the inner region than the distributions around the
quasar positions, and the overdensities within 35'' are about twice more
significant. In fact, there are fields (e.g. 1435+383) where the
smoothness due to the off-centering of the quasar is dramatic (compare
Fig. 2 and Fig. 5). In all the fields (except perhaps 1435+383), the
quasar is within the overdensity range.

In Fig. 5 it is
also seen that the excess still presents a smooth component in some of
the quasars fields, although it is less significant than in Fig. 2.
Therefore, although the off-centering artificially
smooths the radial distribution of relative excess, our
result does not completely exclude
the other two hypothesis to explain the {\it large
scale} or {\it smooth} component.

% Figure 5
   \begin{figure*}
\vspace{0cm}
\hspace{0cm}\epsfxsize=16cm \epsfbox{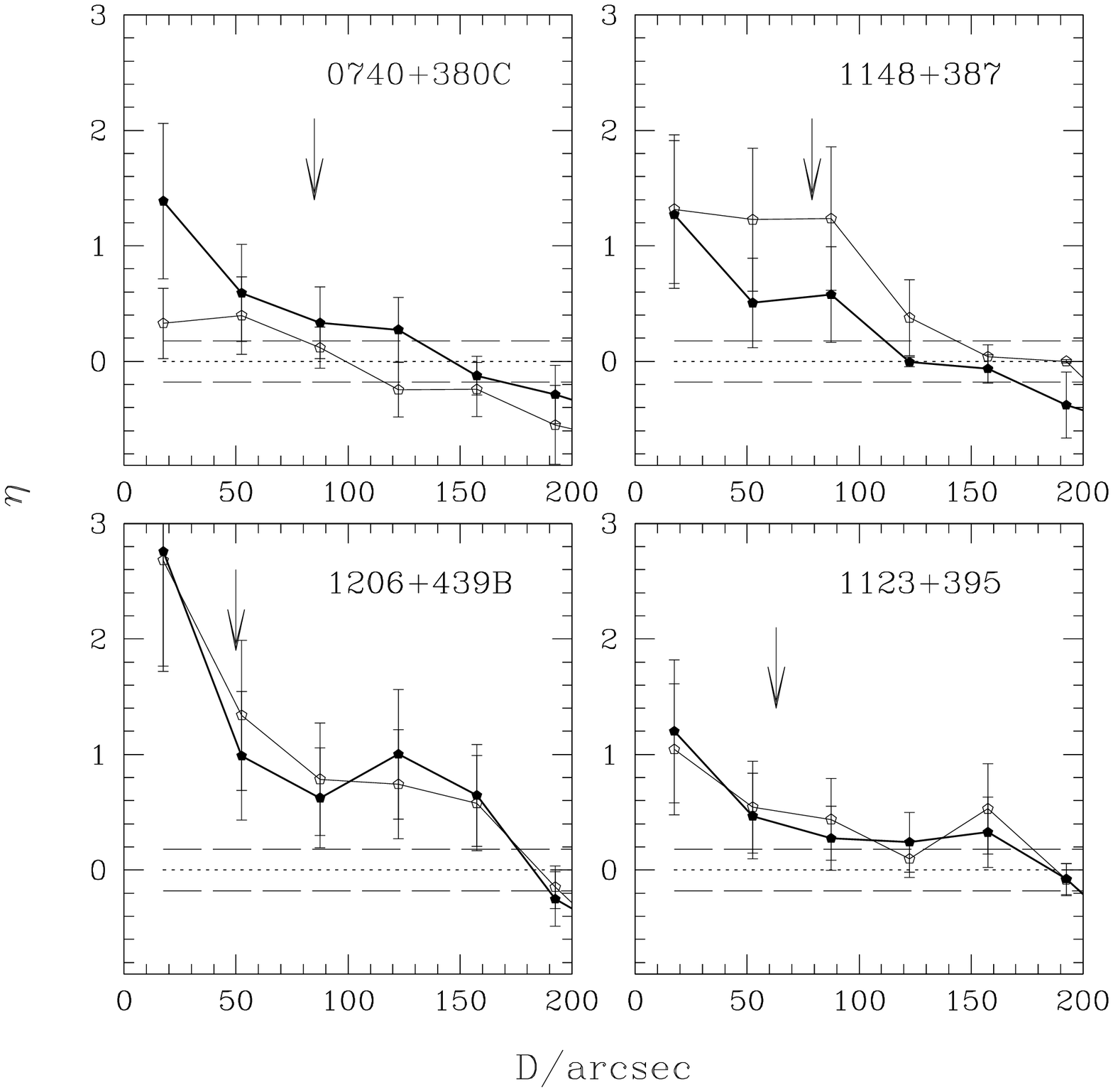}
\vspace{0cm}

\caption{Radial distribution of the relative
 excess of galaxies ($\eta$) with respect to the position of the
 $center$ of the density (in arcsec) for each quasar field and averaged
 over all the fields (indicated as mean). The thick and thin solid
 lines correspond to the objects detected in the $R$-band and $B$-band
 images, respectively . The arrows mark the position of the quasar in
 the field. }
         \label{Figure 4}%
\end{figure*}

\setcounter{figure}{4}
% Figure 5
   \begin{figure*}
\vspace{0cm}
\hspace{0cm}\epsfxsize=16cm \epsfbox{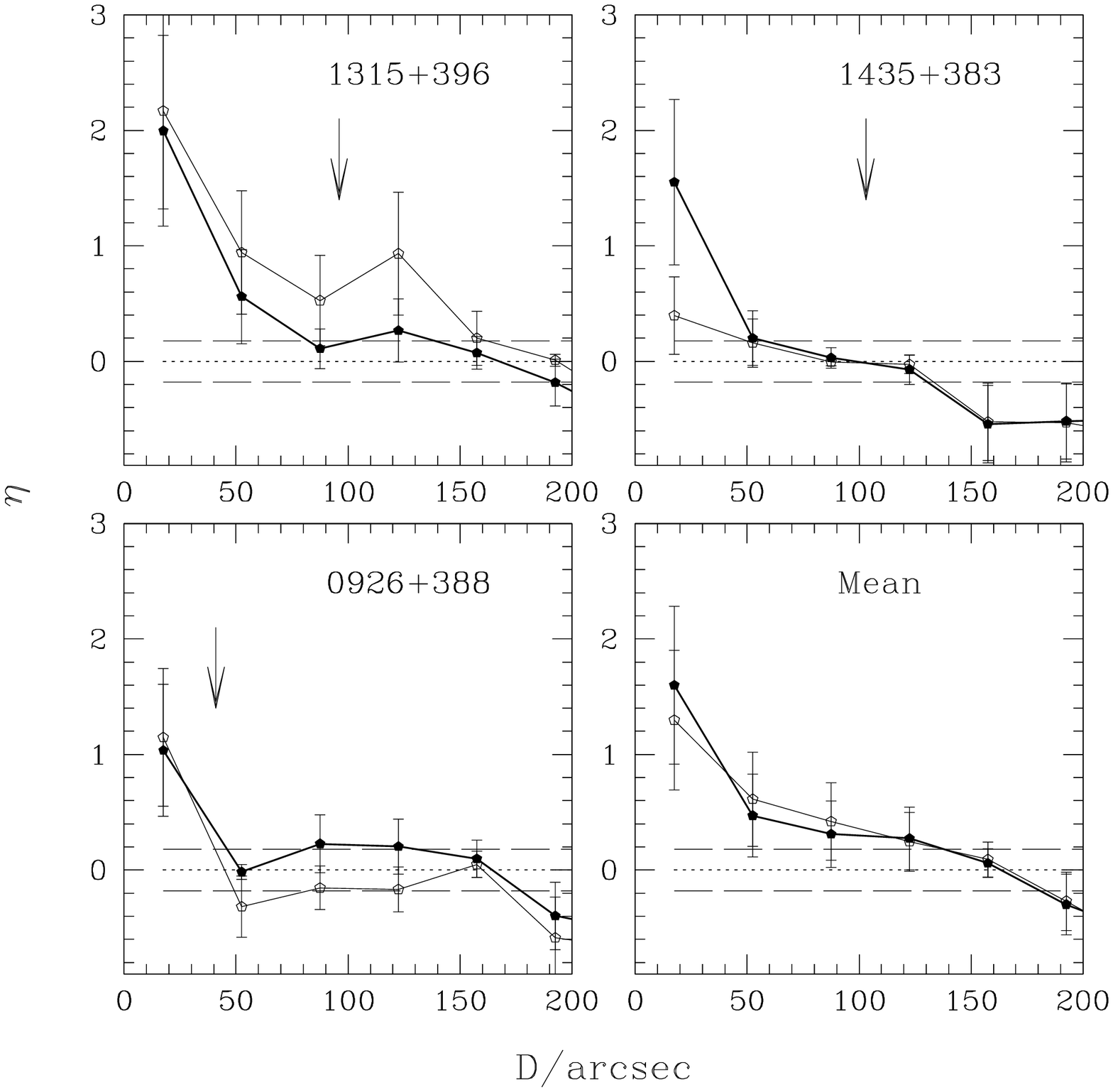}
\vspace{0cm}
\caption{{\it Continued}Radial distribution of the relative
 excess of galaxies ($\eta$) with respect to the position of the
 $center$ of the density (in arcsec) for each quasar field and averaged
 over all the fields (indicated as mean). The thick and thin solid
 lines correspond to the objects detected in the $R$-band and $B$-band
 images, respectively . The arrows mark the position of the quasar in
 the field. }
         \label{Figure 5b}%
\end{figure*}

\section{Summary and conclusions}

In this study, the environments of 7 radio-loud quasars at a redshift
range between $\sim$1 and $\sim$1.6 have been observed in the $B$ and
$R$ band. A significant excess (up to $\sim3\sigma$) of faint galaxies
($R>$22, $B>$22.5) have been found around them, with magnitudes, angular
size of the excess, and number of galaxies all compatible with clusters at
the quasar redshifts. This result agrees with previous results
on radio-loud quasars at similar redshift ranges (Hintzen et al. 1991,
Hall \& Green 1998), and it is in contrast with results reported on
radio-quiet ones (Boyle \& Couch 1993). As at lower redshift ranges,
the environments of radio-loud and radio-quiet sources appear to be
significantly different (\cite{yg87}, \cite{eyg91}).

We have found a tendency between radio power and excess of
galaxies around QSOs.
This trend could be explained by two different
mechanisms: (1) If the intracluster gas plays the role of the fuel of
the nuclear engine, and radio power is a good estimator of its
activity, it is expected that radio power increases with gas
density. It is also expected that the amount of intracluster gas were
larger in high density clusters, which could explain the correlation;
(2) The steep-spectrum extended radio emission is explained by the
interaction of $jets$ with the
intergalactic medium (e.g. Miley 1980). This medium, with a density of
at least $\rho\ge 10^{-28}$ g$\cdot$ cm$^{-3}$, confines the radio
lobes. It is expected that intergalactic medium were more dense in
more populated clusters, and, therefore, there were a connection
between cluster population and radio power. Maybe the combination of
both mechanisms has to be invoked to explain the proposed tendency. It
is interesting to note here that at least the second hypothesis could
explain the connection between excess and slope of radio emission
found by Hall \& Green (1998) and Mendes de Oliveira et al. (1998).

Two possible explanations to the observed link between nuclear
activity and environment have been previously proposed: the dynamical
state of the cluster galaxies, the evolution of intracluster gas
(\cite{yg93}). In both schemes the merger or interaction between the
host galaxy and companions galaxies plays a substantial role, which
reinforces the suggestion that the nuclear activity is triggered by
merging processes (Barnes \& Hernquist 1992, Shlosman 1994). There are
other evidences that support this suggestion, like the large fraction
of host-galaxies undergoing tidal interactions/merging process, their
large absolute magnitudes, and the velocity distribution of galaxies
around quasars (Smith et al. 1986; Hutchings 1987; Disney et al. 1995;
Hutchings \& Neff 1997; Carballo et al. 1998). Another possible
explanation could be a connection between the mass of the central
supermassive black hole and the depth of the potential well of the
cluster, which could be reflected in the richness of the cluster itself.
It is not clear which could be the primary physical cause among the
above refereed, and maybe a combination of all of them would be needed
to explain the results.

We have found that the quasars of our sample are {\it not } located at
the center of the clusters. The off-centering of the quasars partially
explains the smoothing of the radial distribution of the excess, which
apparently shows a two-component distribution (Hall \& Green 1998). A
better determination of the nature of the overdensity is needed before
to present a conclusive explanation for this off-centering. However,
this result is not completly unexpected in the framework of merging as
the triggering/fuelling mechanism of the nuclear activity: Aarseth \&
Fall (1980) have shown that the merging process requires that the
encounter velocity does not significantly exceed the internal velocity
dispersion of the galaxies, approximately 200 km s$^{-1}$. Therefore,
in virialized cluster cores, galaxy-galaxy merging would be less
effective that in the outer part of the clusters, where we have found
that the quasars of our sample inhabit.

Among the follow-up research projects suggested by this work, the most
important is to select a sample of candidates to cluster
members among the galaxies detected in each field. It would be
interesting to obtain optical-NIR multiband imaging and multislit
spectroscopy of these candidates in order to verify the existence of
overdensities at the quasars redshifts, to discriminate the age and
metallicity of the selected galaxies, and to do a dynamical study of
the putative clusters in order to determine why the active nuclei are
not at the center of them.

\begin{acknowledgements}

The 2.2m telescope, at the Centro Astron\'omico
Hispano-Alem\'an, Calar Alto, is operated by the Max-Planck-Institute for
Astronomy, Heidelberg, jointly with the spanish Comisi\'on Nacional de
Astronom\'\i a.

We also thank S. Charlot for providing us with the GISSEL package, and
L. Cay\'on and I. Ferreras for their help in obtaining the $k+e$ and
evolutionary corrections from the GISSEL models for the used
filters. We thank E. Mart\'\i nez-Gonz\'alez for the useful comments
about the derivation of the number-counts variance, and I. Ferreras for
comments about the luminosity of the galaxies. We also thank the
referee for the useful coments. Financial support was
provided under DGICYT project PB92 0501, DGES project PB95 0122, and
by the Comisi\'on Mixta Caja Cantabria - Universidad de
Cantabria. S.F. S\'anchez wants to thank the FPU/FPI program of the
Spanish MEC for provide him a PhD student fellowship.

\end{acknowledgements}

\end{document}